\documentstyle[11pt,newpasp,twoside, epsf]{article}
\def\edcomment#1{\iffalse\marginpar{\raggedright\sl#1\/}\else\relax\fi}
\reversemarginpar

\begin{document}
\title{Publications of the Astronomical Society of the Pacific Conference Series--- \\ Host galaxies, BH masses and Eddington ratio of radio-loud AGNs}
 \author{Aldo Treves and Nicoletta Carangelo }
\affil{Universit\`a dell'Insubria, Via Valleggio 1, Como, Italy}
\author{Renato Falomo}
\affil{Osservatorio Astronomico di Padova, Vicolo dell' Osservatorio 5, Padova, Italy}
\author{C. Megan Urry and Matthew O'Dowd }
\affil{Space Telescope Science Institute, 3700 San Martin Dr., Baltimore, MD 21218, USA}
\author{Riccardo Scarpa}
\affil{European Southern Observatory, Alonso de Cordova 3107, Vitacura, Casilla 19001, Santiago, Chile}

\begin{abstract}

We compare the host galaxies properties of BL Lac objects with those
of radio loud quasars (RLQs) imaged by the 
 WFPC2 on board of HST. 
The considered  objects (z$<$0.5) are always well resolved and their host galaxies satisfactorily
modelled by ellipticals.  After homogeneous treatment of the data we
find RLQs hosts are systematically more luminous (by $\sim$ 0.7 mag)
with respect to the hosts of BL Lacs. Using the M$_{BH}$ - L$_{bulge}$ relation, derived for nearby elliptical galaxies, we have evaluated the
central black hole masses of our sample of active galaxies.  These data are discussed in
conjunction with the nuclear luminosity and the Eddington ratio.

\end{abstract}

\section{Introduction}
A clear insight
of the galaxies hosting active nuclei is a fundamental ingredient for
the understanding of both galaxies and nuclei formation and
evolution. A comparison of the host galaxies of different type of AGN can
for instance provide a direct test for unification models (e.g. Urry and Padovani 1995 ) while the
relationship between nuclear properties and the  immediate environment 
yields clues for the link between the formation of galaxies and the growth 
and evolution of a massive black hole ( e.g. Kauffmann \& Haehnelt 2000).\\ 

A good characterization of the host galaxies properties requires
images of excellent quality in order to disentangle the light of the
galaxy from that of the bright nucleus.  To this aim HST images have
provided a significant improvement of data on AGN host galaxies at
various redshifts (Disney et al. 1995, Bahcall et al. 1996, 1997,
Boyce et al. 1998, McLure et al. 1999; Hamilton 2000, Scarpa et al. 2000, Urry et al. 2000, Kukula et al. 2001).\\

In this paper we have considered two samples of active
galaxies imaged by HST  including respectively BL Lacs and RLQs. We compare and discuss
their host properties in conjunction with the nuclear luminosity,
estimated black hole (BH) masses and Eddington ratio.
Preliminary results on the arguments discussed here were presented by 
O'Dowd et al. (2001).

\section{The samples}

We have collected data for all BL Lacs and RLQs at z$<$0.5 observed
with the WFPC2 of HST. These observations were in general secured in different
filters and were analyzed using different methods and calibrations.
Our analysis has provided a uniform and homogeneous dataset of host
galaxies for low z radio loud active galaxies.  Most of the observations 
were obtained in the F702W filter therefore we converted all
magnitudes into R (Cousins) band. Color corrections for objects
observed in other filters were derived using the expected colors of
galaxies given by Fukugita et al. (1995). Absolute
magnitudes have been computed assuming for H$_{0}$=50 Km s$^{-1}$
Mpc$^{-1}$ and of $\Omega_{0}$=0.  In addition K-correction was
performed following Poggianti (1997) and
galactic reddening corrections as
 in Scarpa et al. 2000.\\

\subsection{The BL Lac sample}

The HST snapshot imaging survey of BL Lacs (Urry et al.  2000, Scarpa et
al.  2000) has provided an  homogeneous set of 110 short
exposure high resolution images through F702W filter. 
From this dataset we have considered all objects at z$<$ 0.5 that are
resolved by the HST image. This yields 57 sources with z between 0.027 and 0.495 and $<z>$=0.20$\pm$0.11. Only for three sources at z$<$0.5 the host has not been detected by HST images (see Scarpa et al.). For these 57  objects  the associated host galaxy morphology is always well described by elliptical modelling.

\subsection{The RLQ sample}

There is not a comparable large set of HST observations for higher
luminosity RLQ therefore we have constructed a representative dataset
for RLQ from collection of various sources reporting RLQ images
secured by HST. In order to provide a homogeneous treatment we
focussed  on data for which apparent magnitudes of the host
galaxies were given, so that the calibration
procedure could be reproduced.  This translates into the collection of RLQs
investigated by Bahacall et al. (1997, 1999) (8 objects), Boyce et al. (1998) (6 objects) and McLure et al. (1999) (6 objects).\\

As in the case of BL Lacs for these objects the host galaxy is always
visible and well consistent with an elliptical morphology.  In
table 1 we report the apparent and absolute magnitudes of the host
galaxies. Our
evaluation of M$_{R}$ is consistent with absolute values reported by the quoted
authors when galactic extinction and filter transformations is taken into account. However in the case of Boyce et al. subsample a larger difference is noted due likely to a mistreating of k-correction  and cosmological transformations.\\

Our comparison of three subsamples of RLQ show they are indistinguishable, therefore in the following analysis we considered 
a unique sample of 15 objects, with redshift in the range 0.158$<z<$0.389 and $<z>$=0.26$\pm$0.07, taking averages values for objects observed twice.\\ 

\begin{quote}
\begin{table}
\centering
\begin{tabular}{|l|l|l|l|l|l|l|l|l|} \hline
Source 	     &    z  & Filt  & m$_{host}$ & R$_{host}$ & M(R)$_{host}$  & A$_{r}$ & Kcorr & Refs. \\ \hline
3C48         & 0.367 & f555w & 17.7 & 17.2 & -25.2	& 0.26 & 0.48 & B \\ \hline 
PHL 1093     & 0.258 & f675w &  *   & 17.2  & -24.4	& 0.14 & 0.3  & McL \\ \hline 
PHL 1093     & 0.258 & f702w & 17.8 & 17.2 & -24.4	&  *   & 0.3  & By \\ \hline 
PKS 0202-76  & 0.389 & f702w & 20.0 & 19.5 & -24.2	& 0.94 & 0.52 & By \\ \hline 
0312-77      & 0.223 & f702w & 17.7 & 17.1 & -24.8	& 0.76 & 0.26 & By \\ \hline 
0736+017     & 0.191 & f675w & *    & 16.9  & -24.3	& 0.46 & 0.21 & McL\\ \hline 
1004+130     & 0.240 & f606w & 16.9 & 16.9 & -24.4	& 0.16 & 0.28 & B \\ \hline 
1004+130     & 0.240 & f675w & *    & 16.9  & -24.6	&  *   & 0.28 & McL\\ \hline 
3C273        & 0.158 & f606w & 16.0 & 15.8 & -24.4	& 0.07 & 0.17 & B \\ \hline 
1302-102     & 0.286 & f606w & 18.2 & 18.2 & -23.4	& 0.09 & 0.34 & B \\ \hline 
1302-102     & 0.286 & f702w & 18.0 & 17.4 & -24.4	&  *   & 0.34 & By \\ \hline 
1425+267     & 0.366 & f555w & 18.3 & 17.8 & -24.4	& 0.09 & 0.48 & B \\ \hline 
3C323.1      & 0.264 & f606w & 18.1 & 18.1 & -23.4	& 0.19 & 0.31 & B \\ \hline 
3C351        & 0.371 & f702w & 18.5 & 18.0    & -24.7	& 0.09 & 0.49 & By \\ \hline 
2135-147     & 0.200 & f606w & 17.4 & 17.4 & -23.7 	& 0.34 & 0.23 & B \\ \hline 
OX169        & 0.213 & f675w & *    & 17.2  & -24.1	& 0.36 & 0.24 & McL\\ \hline 
2247+14      & 0.237 & f675w & *    & 17.2  & -24.3	& 0.24 & 0.27 & McL\\ \hline 
2349-014     & 0.173 & f606w & 16.2 & 16.0 & -24.5	& 0.16 & 0.19 & B \\ \hline 
2349-014     & 0.173 & f675w & *    & 15.9  & -24.7	& *    & 0.19 & McL\\ \hline 
\end{tabular}
\caption{Sample of RLQs: we report in column (1) the source, (2) the redshift, (3) the observational filter F, (4) the apparent magnitude in filter F, (5) the apparent magnitude of host galaxy in Cousins filter R, (6) the absolute magnitude of host galaxy in Cousins filter R, (7) the extinction correction in filter R, (8) the K-correction and in column (9) the reference (B=Bahcall et al. 1997; By= Boyce et al. 1998; McL McLure et al. 1999).}
\end{table}
\end{quote}

\begin{quote}
\begin{figure}  
\plotfiddle{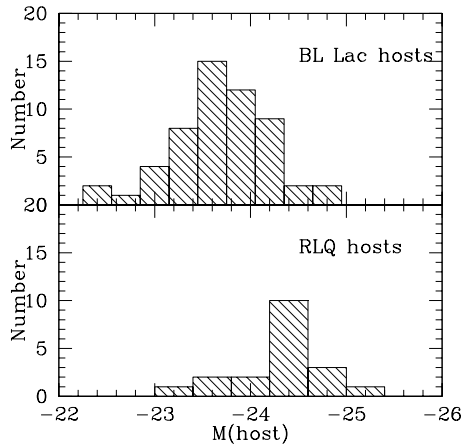}{5 cm}{0}{30}{30}{-100}{-45}
\caption{Absolute R magnitude distribution of BL Lac (upper pannel) and RLQ (lower pannel) host galaxies : RLQ hosts are on average $\sim$0.7 mag more luminous than those of BL Lacs.}
\end{figure}
\end{quote}

\section {Results and discussion}

\subsection{Comparison of host galaxies of BL Lacs and RLQs}

The comparison of the distribution of host galaxies luminosities (see
Fig 1) shows that RLQ hosts are on average more luminous than those of
BL Lacs. This should not be an effect due to selection of objects since 
both QSO and BL Lacs were chosen basing only on the nuclear properties.\\

The average host absolute magnitude for the BL Lacs is $<M_{R}>_{BLL}$
= --23.7 $\pm$ 0.52; while for RLQ hosts we find 
$<M_{R}>_{RLQ}$ = --24.32 $\pm$ 0.44. This systematic difference is not
due to the different redshift range covered by the two sample
since if we use only BL Lacs in the range 0.15$<z<$0.4 to
match the one of the RLQ sample we find $<M_{R}>_{BLL}$ = --23.77 $\pm$
0.49.\\

In the recent analysis of a large sample of low z ($<$ 0.46) QSOs
 Hamilton et al. 2001 report absolute V mag for 26  RLQs. After
 conversion to our cosmology, including correction for galactic
 extinction and assuming V-R=0.6 the average value of the host luminosity is $<M_{R}>_{RLQ}$ = --24.8, leading to an even larger difference with BL Lac hosts.

\section{The masses of the central Black Hole}

Dynamical studies of nearby  elliptical galaxies have shown that there is 
a correlation between the galaxy luminosity and the 
mass M$_{BH}$ of the central
black hole (Magorrian et al. 1998; Kormendy and Gebhardt 2001 and references
therein). A much tighter correlation was found between the BH mass
and the stellar velocity dispersion of the host bulge of the galaxy
(Ferrarese \& Merritt 2000; Gebhardt et al. 2000). Moreover reverberation mapping estimates (Kaspi et al. 2000)
of M$_{BH}$ of nearby AGN seem to indicate that this correlation holds
also for massive ellipticals hosting an active nucleus.  With this
assumption it is therefore possible to estimate the
BH mass in AGN from the integrated host luminosity and/or the velocity
dispersion (which is much more difficult to measure).\\

In figure 2, adapted from Kormendy and Gebhard 2001, 
the M$_{BH}$ -- M$_{B,bulge}$ is reproduced together with the
regions covered by the estimated BH masses for our two samples of BL Lacs and
RLQs. It turns out that BH masses are in the range
5$\times$10$^{8}$ to 5$\times$10$^{9}$ and cover the region of most massive objects. 
It also turns out that on average BH masses of RLQ are a factor $\sim$ 2 
greater than those of BL Lacs as a direct consequence of the different 
average luminosity of their hosts ($<M_{bh}>$=1.5$\times$ 10$^{9}M_{\sun}$ for BL Lacs and
$<M_{bh}>$=2.4$\times$ 10$^{9}M_{\sun}$ for RLQs). 

\begin{quote}
\begin{figure}  
\plotfiddle{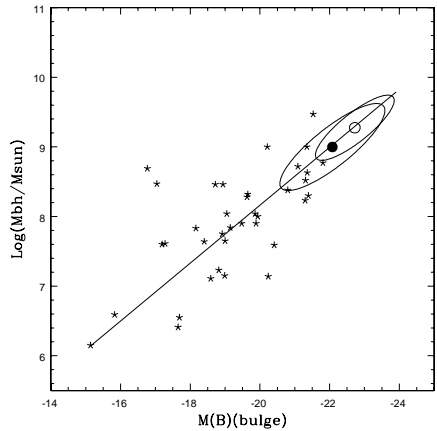}{5 cm}{0}{30}{30}{-100}{-50}
\caption{Relation between the black hole mass and the absolute magnitude of the bulge component of the host galaxy for a sample of 37 nearby galaxies and linear fit (Kormendy \& Gebhardt 2001). The filled and empty circle represent respectively the position corresponding to average values of black hole masses for BL Lacs and RLQs. The ellipses indicate the ranges covered by the two classes of objects}
\end{figure}
\end{quote}

\section {Eddington ratios} 

The estimated BH masses can be used to derive the Eddington luminosity L$_{E}$ and can be compared with the total luminosity L deduced by the spectral energy distribution (SED), yielding the Eddington ratio $\xi_{E}=L/L_{E}$.\\

For RLQs we assume an average total luminosity of L$\simeq3\times$
10$^{12} L_{\sun}$ (Elvis et al. 1994). In the case of the BL Lacs, in order to evaluate the
intrinsic emitted total luminosity one has to take into account 
 the beaming. We took L$\propto$F$\times\delta^{-2}$, where F is the observed flux
and $\delta$ is the Doppler factor. For the SED of a large
sample of BL Lacs we refer to Sambruna et al.  (1996) and 
Fossati et al. (1998). The average beaming factor is taken $\delta\sim$ 15 as suggested 
by Ghisellini et al.  (1998) (see also Capetti and Celotti 1998). The resulting average intrinsic total luminosity for BL Lacs is therefore 
L$\simeq$ 10$^{10}$ L$_{\sun}$.\\

By the above arguments the BH masses of the two samples are similar
(within a  factor of 2) while the intrinsic  luminosities
of the two classes differ by a factor 
$\sim$ 100.
The main inference is thus that {\it the Eddington ratio $\xi_{E}$, differs by two orders of magnitude in the two classes}.

\begin{quote}

\end{quote}

\end{document}